\documentclass[twocolumn,showpacs,preprintnumbers,amsmath,amssymb]{revtex4}
\input psfig.sty
\def \be {\begin{equation}}

\def \ee {\end{equation}}
\def \bea {\begin{eqnarray}}
\def \eea {\end{eqnarray}}

\usepackage{graphicx}
\usepackage{dcolumn}
\usepackage{bm}

\begin{document}

\title{Constraining the dark energy and smoothness-parameter with supernovae}

\author{R. C. Santos} \email{cliviars@astro.iag.usp.br}

\author{J. V. Cunha} \email{cunhajv@astro.iag.usp.br}

\author{J. A. S. Lima} \email{limajas@astro.iag.usp.br}
\vskip 0.5cm
\affiliation{Departamento de Astronomia, Universidade de S\~ao Paulo, 05508-900 S\~ao
Paulo, SP, Brasil}

\pacs{Dark energy, cosmic distance, supernovas, inhomogeneities}
\begin{abstract}
The presence of inhomogeneities modifies the cosmic distances
through the gravitational lensing effect, and, indirectly, must affect the main cosmological tests. Assuming that the dark energy is a smooth component, the simplest way to account for the influence of clustering is to suppose that the average evolution of the expanding Universe is governed by the total matter-energy density whereas the focusing of light is only affected by a fraction of the total matter density quantified by the $\alpha$ Dyer-Roeder parameter.  By using two different samples of SNe type Ia data, the  $\Omega_m$ and $\alpha$ parameters are constrained by applying the Zeldovich-Kantowski-Dyer-Roeder (ZKDR) luminosity distance redshift relation for a flat ($\Lambda$CDM) model. A $\chi^{2}$-analysis using the 115 SNe Ia data of  Astier {\it et al.} sample (2006) constrains the density parameter to be  $\Omega_m=0.26_{-0.07}^{+0.17}$($2\sigma$) while  the $\alpha$ parameter is weakly limited (all the values $\in  [0,1]$ are allowed even at 1$\sigma$). However, a similar analysis based the 182 SNe Ia data of Riess {\it et al.} (2007) constrains the pair of parameters to be $\Omega_m= 0.33^{+0.09}_{-0.07}$ and $\alpha\geq 0.42$ ($2\sigma$). Basically, this occurs because the Riess {\it et al.} sample extends to appreciably higher redshifts.  As a general result, even considering the existence of inhomogeneities as described by the smoothness $\alpha$ parameter, the Einstein-de Sitter model is ruled out by the two samples with a high degree of statistical confidence ($11.5\sigma$ and $9.9\sigma$, respectively). The inhomogeneous Hubble-Sandage diagram discussed here highlight the necessity of the dark energy, and a transition deceleration/accelerating phase at $z\sim 0.5$ is also required.
\end{abstract}

\maketitle

\section{Introduction}
The Hubble-Sandage diagram for Type Ia Supernovae (hereafter SNeIa), as
measured by the Supernova Cosmology Project\cite{perm98} and
the High-z Supernova Search Team\cite{Riess}, provided the first evidence that the
present Universe is undergoing a phase of accelerating expansion driven by an exotic component with negative pressure (in addition to the cold dark matter), usually called dark energy. 

The idea of a dark energy-dominated
universe is a direct consequence of a convergence of independent
observational results, and constitutes one of the greatest
challenges for our current understanding of fundamental physics\cite{review}. Among a number of possibilities to describe this
dark energy component, the simplest and most theoretically
appealing way is by means of a cosmological constant $\Lambda$,
which acts on the Einstein field equations as an isotropic and
homogeneous source with a constant equation of state, $w
\equiv p/\rho = -1$. 
 
Although cosmological scenarios with a $\Lambda$ term might explain most of the current astronomical observations, from the theoretical viewpoint they are plagued with at least a fundamental problem, namely, it is really
difficult to reconcile the small value of the vacuum energy density required by observations ($\simeq 10^{-10} \rm{erg/cm^{3}}$) with estimates from quantum field theories ranging from 50-120 orders of magnitude larger\cite{weinberg}. This problem sometimes called the cosmological constant problem (PCC) has inspired many authors to propose decaying $\Lambda$ models\cite{list} and other alternative approaches for describing dark energy\cite{list1}. Nevertheless, the present cosmic  concordance model (CCM) which is supported by  all the existing observations is a flat $\Lambda$CDM cosmology with a matter fraction of $\Omega_{\textrm{m}} \sim 0.26$ and a vacuum energy contribution of $\Omega_{\Lambda}\sim 0.74$\cite{Astier06,Sperg07,Ries07,Allen07}. 

On the other hand, the
real Universe is not perfectly homogeneous, with light beams
experiencing mass inhomogeneities along their way thereby producing many  observable phenomena. For instance, light lines traversing in the universe are attracted and refracted by the gravitational force of the galaxies on their path, which bring
us the signal of lensing, one of which is the multiple
images of a single far galaxy\cite{SFE92,Ogu07}. Nowadays, gravitationally lensed quasars and radio sources offer important
probes of cosmology and the structure of galaxies. The optical depth
for lensing depends on the cosmological volume element out to
moderately high redshift. In this way, lens statistics can in principle
provide valuable constraints on the cosmological constant or, more
generally, on the dark energy density and its equation of state
\cite{estatist,Kasai,Schneider}.

In this context, one of the most important issues in the modern cosmology is to quantify  from the present observations the influence of such inhomogeneities on the evolution of the Universe. An interesting possibility to account for such effects is to introduce the
smoothness parameter $\alpha$ which represents the magnification
effects experienced by the light beam. When $\alpha = 1$ (filled
beam), the FRW case is fully recovered; $\alpha < 1$ stands for a
defocusing effect; $\alpha = 0 $ represents a totally clumped
universe (empty beam). The distance relation that takes the mass
inhomogeneities into account is usually named Dyer-Roeder
distance\cite{Dy72}, although its theoretical necessity had been previously studied by
Zeldovich\cite{Ze64} and Kantowski\cite{Kant69}. In this way, we label it here 
as Zeldovich-Kantowski-Dyer-Roeder (ZKDR) distance formula (for an overview on cosmic distances taking into account the presence of inhomogeneities see the paper by Kantowski\cite{Kant03}). 

Several studies involving the ZKDR
distances in dark energy models have been published in the last few years. Useful analytical expressions for $\Lambda$CDM models have been derived by Kantowski {\it et al.} \cite{Kant98,Kant00} and Demianski {\it et al.}\cite{DEM03}. Working in the empty beam approximation ($\alpha = 0$), Sereno {\it et al.}\cite{SPS01} investigated  some effects of the ZKDR distance for a general background. By assuming that both dominant components may be clustered they also discussed  the critical redhift, i.e.,
the value of $z$ for which $d_{A}(z)$ is a maximum (or $\Theta(z)$
minimum), and compared to the homogeneous background results as
given by Lima and Alcaniz\cite{ALDA00}, and further discussed by Lewis
and Ibata\cite{LIB02}.  Demianski and coworkers derived an approximate solution for a clumped concordance model valid on the interval $0\leq z \leq 10$. Additional studies on this subject is related to time delays\cite{GA01,LIB02}, gravitational
lensing\cite{koc02,koc03}, and even accelerated models driven by particle creation have been investigated\cite{CdS04}. 

In a previous paper\cite{SL07}, we have applied the ZKDR equation in the framework of phantom cosmology in order to determine cosmological constraints from a sample of milliarcsecond compact radio sources. By assuming a Gaussian prior on the matter density parameter, i.e.,
$\Omega_m = 0.3 \pm 0.1$, the best fit model for a phantom cosmology
with $\omega = -1.2$ occurs at $\Omega_m = 0.29$ and $\alpha = 0.9$
when we marginalize over the characteristic size of the compact
radio sources. Such results suggested that the ZKDR distance can give important corrections to the so-called background tests of dark energy.
\begin{figure*}
\centerline{\psfig{figure=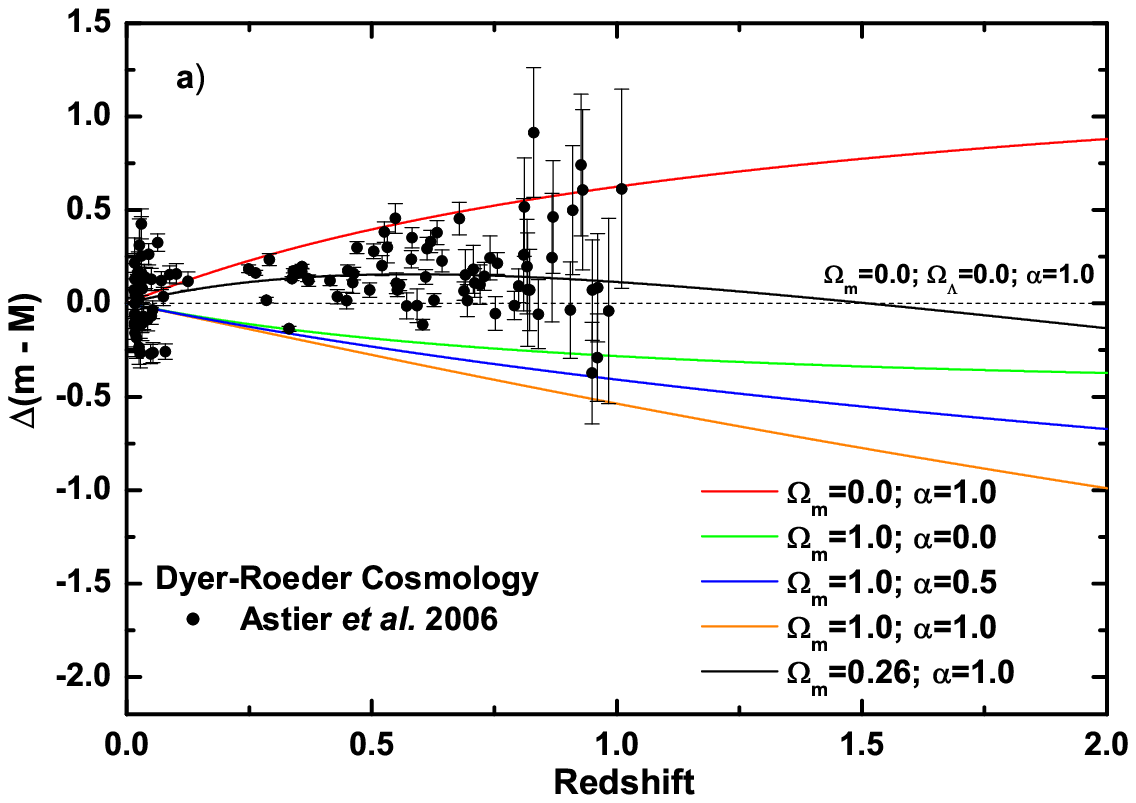,width=3.1truein,height=2.5truein}
\psfig{figure=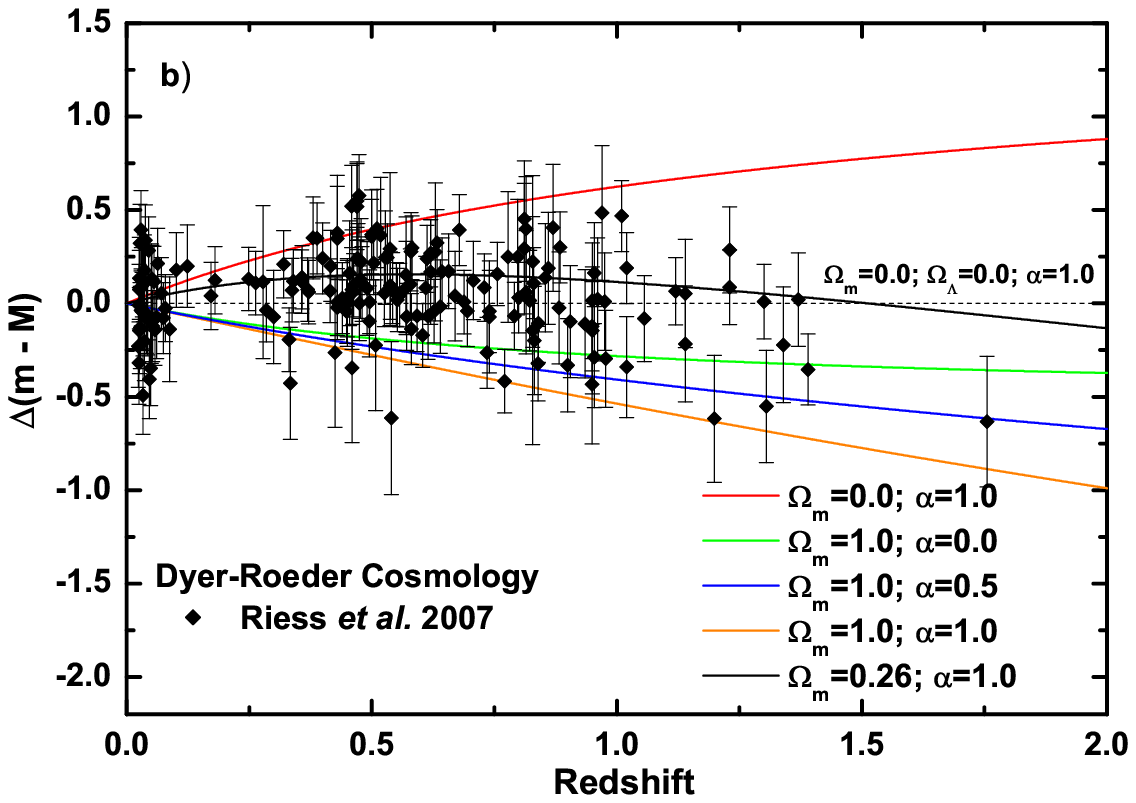,width=3.1truein,height=2.5truein}
\hskip 0.1in} \caption{The $\alpha$-effect on the residual magnitudes. In Fig. {\bf a} we show the 115 Supernovae data from Astier  {\it{et al.}}\cite{Astier06}, and the predictions of the ZKDR luminosity distance for several values of $\alpha$ relative to an empty model ($\Omega_m=0$, $\Omega_{\Lambda}=0$ and $\alpha=1$). In Fig. {\bf b} we show the same graph but now for the  182 SNe type Ia from Riess {\it{et al.}} sample \cite{Ries07}. For comparison, in both panels we see (black curves) the prediction of the cosmic concordance model ($\Omega_m=0.26$, $\Omega_{\Lambda}=0.74$, $\alpha=1$).}
\end{figure*}
In this article, the pair of cosmic  parameters, $\Omega_m \equiv 1-\Omega_{\Lambda}$ and $\alpha$, are constrained from Supernovae observations by applying the ZKDR distance-redshift relation for a flat ($\Lambda$CDM) model.  As we shall see, the $\alpha$ parameter is not well
constrained by the 115 SNe observed by Astier {\it et al.} (2006). However, the 182 SNe Type Ia sample of Riess {\it et al.} (2007), constrains the pair of parameters to be $\Omega_m= 0.33^{+0.09}_{-0.07}$ and $\alpha\geq 0.42$ ($2\sigma$). As a general result, even considering the existence of inhomogeneities described by the $\alpha$ parameter, the Einstein-de Sitter model is ruled out by the two samples with a high degree of statistical confidence ($11.5\sigma$ and $9.9\sigma$, respectively). 

The paper is organized as follows. In section II, we present the basic
equations and the distance description taking into account the  inhomogeneities 
as described by the ZKDR equation. In section III,
we determine the constraints on the cosmic parameters from the two Supernovae samples. Finally, we summarize the main conclusions in section IV.

\section{ZKDR Equation for luminosity distance}

In a clumpy universe model, the local geometry is inhomogeneous,
but its global aspect can be described by the FRW type geometry ($c=1$)
\begin{equation}
ds^2 = dt^2 - R^2 (t)( dr^2 + r^2 d\Omega^2),
\end{equation}
where $R(t)$ is the scale factor and $d \Omega^2$ denotes the metric in the 2-sphere. 

As it is widely known, the idea of clumpy universe is still a ill-defined notion since we do not have a clear mathematical recipe to separate the global properties from the local inhomogeneous aspects of the Universe. After Dyer \& Roeder\cite{Dy72}, it is usual to introduce a phenomenological parameter, $\alpha=1-{\rho_{cl}\over <\rho_m>}$, called the ``smoothness" parameter. Such a parameter quantifies the
portion of matter in clumps ($\rho_{cl}$) relative to the amount
of background matter which is uniformly distributed ($\rho_m$). In general, due to the structure formation process, it should be dependent of the redshift, as well as, on the direction along the line of sight (see, for instance, \cite{Kasai,SL07} and Refs. there in). However, in the majority  of the works $\alpha$ is assumed to be a constant parameter. From a mathematical viewpoint the treatment is based on the optical-scalar equation for light propagation in the so-called geometric optics approximation\cite{SFE92,Sachs61}
\begin{eqnarray}\label{sachs}
{\sqrt{A}}'' +\frac{1}{2}R_{\mu \nu}k^{\mu}k^{\nu} \sqrt{A}=0,
\end{eqnarray}
where a prime denotes differentiation with respect to the affine
parameter $\lambda$, $A$ is the cross-sectional area of the light
beam, $R_{\mu\nu}$ the Ricci tensor, and $k^{\mu}$ the photon
four-momentum. In this form, it is implicit that the influence of the Weyl tensor
(shear) can be neglected. This means that the light rays are propagating far from the  mass inhomogeneities  so that the large-scale homogeneity implies that their
shear contribution are canceled. The proportionality factor between the cross-sectional length $A^{\frac{1}{2}}$ and the angular distance
$d_A$ can be defined to be constant. Actually, the above optical scalar equation is usually written in terms of the dimensionless angular diameter distance $D_A = H_0 d_A$. Further, by recalling the existence of a simple
relation between the luminosity distance, and the
angular-diameter distance (from Etherington principle\cite{ETHER33},
$d_L = (1+z)^2 d_A$), it is easy to show that the ZKDR (dimensionless) luminosity
distance for $\Lambda$CDM cosmology satisfies the following differential equation\cite{Kant98,Kant00,DEM03,SPS01,GA01}

\begin{equation}\label{angdiamalpha}
 \left( 1+z\right) ^{2}{\cal{F}}
\frac{d^{2}D_L}{dz^{2}} - \left( 1+z\right) {\cal{G}}
\frac{dD_L}{dz} + {\cal{H}} D_L=0,
\end{equation}
which satisfies the boundary conditions:
\begin{equation}
\left\{
\begin{array}{c}
D_L\left( 0\right) =0, \\
\\
\frac{dD_L}{dz}|_{0}=1.
\end{array}
\right.
\end{equation}
where $\cal{F}$, $\cal{G}$ end $\cal{H}$ are functions of the cosmological parameters, expressed in terms of the redshift by:

\begin{eqnarray}
{\cal{F}}& =& \Omega_m + (1-\Omega_m )(1+z)^{-3},\nonumber
\\ \nonumber \\ {\cal{G}} &=& \frac{\Omega_m}{2}
+2(1-\Omega_m )(1+z)^{-3},\nonumber
\\ \nonumber
\\ {\cal{H}} &=& \left(\frac{3\alpha-2}{2}\right)\Omega_m 
 + 2(1-\Omega_m)(1+z)^{-3}, \\
\nonumber
\end{eqnarray}
where as remarked before, the $\alpha$ parameter appearing in the $\cal{H}$ expression (here assumed to be a constant) quantifies the clustered fraction of the pressureless matter. 
                                                                                         
\section{Samples and Results}

The standard FRW models contain only homogeneously and isotropically
distributed perfect fluid gravity sources, and the present CCM is assumed to 
represent both the ``large scale" geometry of the universe and the matter content. 
However, the Universe appears homogeneous only in a statistical sense, when one is describing  the largest scales.  Therefore,  although making very useful predictions, our cosmological models are somewhat inadequate at small and moderate scales. This means that relations like $\mu(H_0,\Omega_m,\Lambda;z)$, the distance modulus for
a standard candle, commonly assumed to be valid on
average can be incorrect even for observations including SNe Ia. In particular if the
underlying mass density approximately follows luminous matter (i.e.,
associated with bounded galaxies),  the effects of inhomogeneities on
relations like $\mu(H_0,\Omega_m,\Lambda,z)$ must be taken into
account. 
\begin{figure*}
\centerline{\psfig{figure=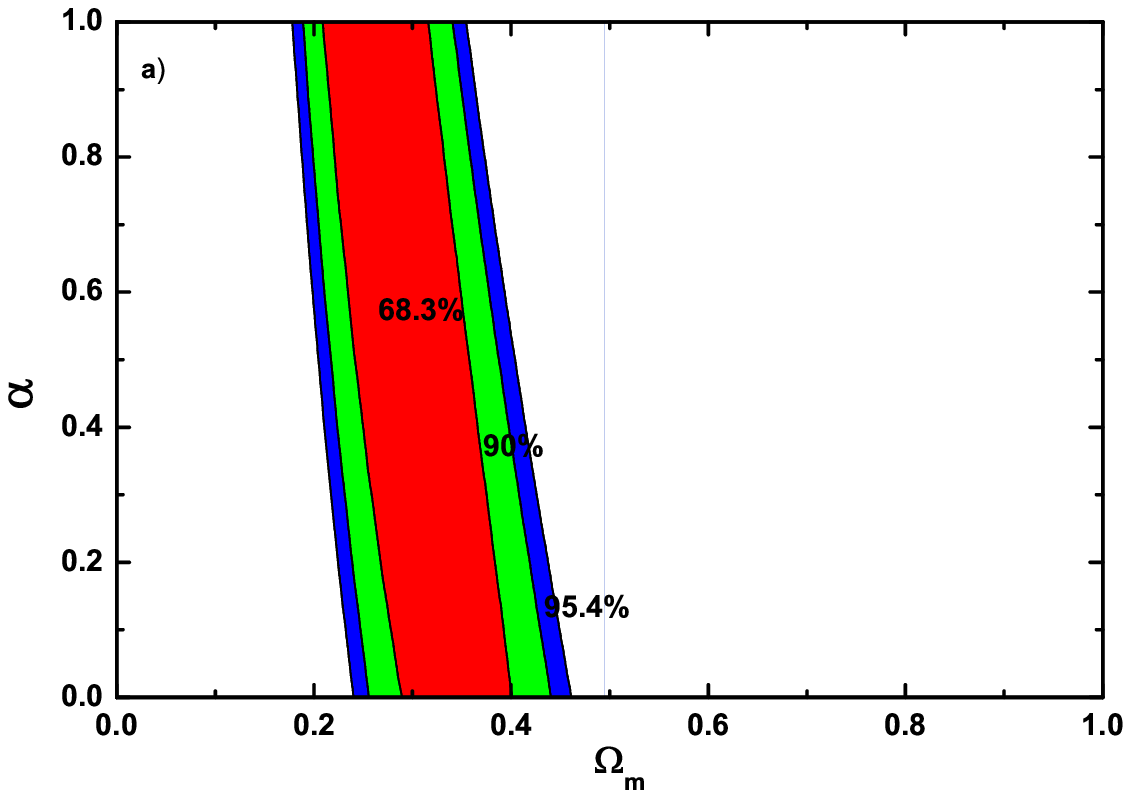,width=2.5truein,height=2.4truein}
\psfig{figure=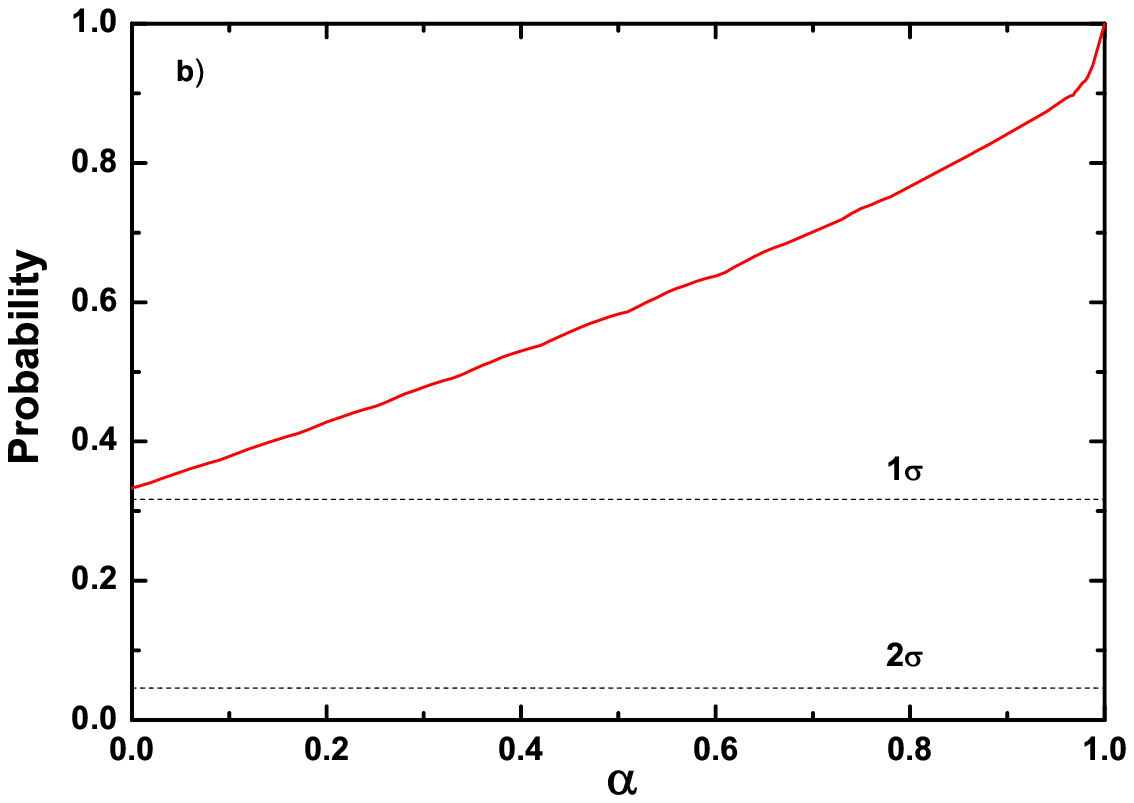,width=2.5truein,height=2.4truein}
\psfig{figure=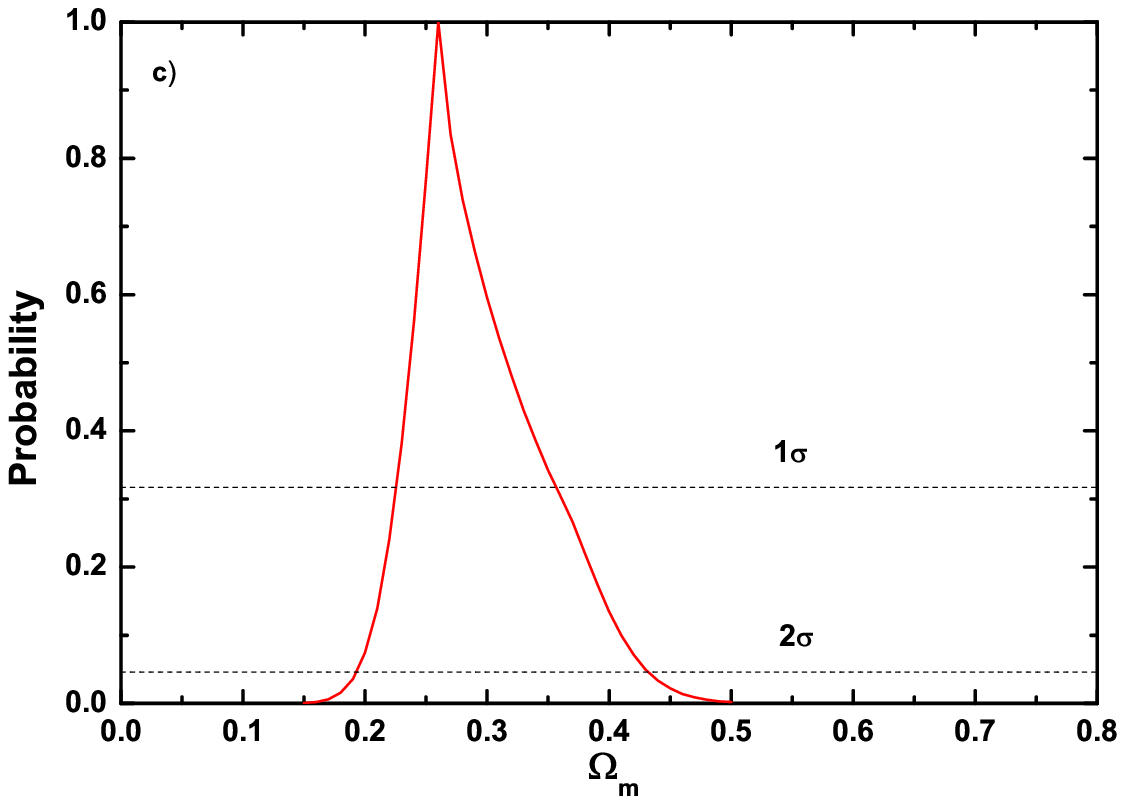,width=2.5truein,height=2.4truein}\hskip
0.1in} \caption{{\bf{a)}} The $\Omega_{m}-\alpha $ plane for flat
$\Lambda$CDM models obtained from 115 SNe Ia data Astier {\it{et
al.}}\cite{Astier06}. Note that the $\alpha$ parameter is not well
constrained by the data. {\bf{b)}} The Likelihood
for the $\alpha$ smoothness parameter. We see that even at 1$\sigma$ the 
smoothness parameter is poorly  restricted (all its admissible values are allowed). {\bf{c)}}
Probability of the matter density parameter. We see that $0.19 \leq \Omega_m \leq 0.43$  
with  $2\sigma$ confidence level.}
\end{figure*}

In Fig. 1, we display the effects of the inhomogeneities in the reduced Hubble-Sandage Diagram for the  Astier {\it et al.} (2006) and Riees {\it et al.} (2007) samples for some selected values of the smoothness parameter. The plots correspond to several
values of $\Omega_m$ and $\alpha$ as indicated in the
panels. The difference between
the data and models from an empty universe case (OCDM) prediction
is also displayed there. For the sake of comparison, we also show the Einstein-de Sitter (E-dS) model, i.e. $\Omega_m=1$ and $\alpha = 1$, as well as the present cosmic concordance ($\Omega_m=0.26$, $\Omega_{\Lambda}=0.74$, $\alpha=1$). Note that the $\alpha$ parameter contributes in the right direction i.e., the SNe type Ia become dimmer when it increases on the allowed range.   
In what follows, a $\chi^{2}$ minimization will be applied for the two sets of SNe data
with the parameters $\Omega_{\rm{m}}$ and $\alpha$ spanning
the interval [0,1] in steps of 0.01, for all numerical computations. 

\subsection{Astier {\it et al}. Sample (2006)}

Let us now discuss the bounds  arising from SNe Ia observations
on the  pair of parameters ($\Omega_{\rm{m}}, \alpha$) defining the ZKDR luminosity distance. 
 
The current data from Supernova Legacy Survey (SNLS) collaboration correspond to the
first year results of its planned five year survey. The total sample
includes 71 high-$z$ SNe Ia in the redshift range $0.2 < z < 1$ plus
44 low-$z$ SNe Ia as published by Astier {\it et al.}\cite{Astier06}. Although in a better agreement with WMAP 3-years results \cite{Sperg07} than the
\emph{gold} sample \cite{Ries07} (for a more detailed discussion see e.g., Jassal
{\it{et al.}}\cite{JS06}), the most distant SNe Ia of these 115 events has redshift smaller than unity. 
 
Following standard lines, the maximum likelihood estimator, ${{\cal{L}}_{SNIa}} \propto \exp\left[-\chi_{SNIa}^{2}(z;\mathbf{p})/2\right]$, is determined by a $\chi^2$ statistics
\begin{equation}
\chi^2_{SNIa}(z|\mathbf{p}) = \sum_i \frac{(\mu (z_i;
\mathbf{p})-\mu_{0,i})^2}{\sigma_{\mu_{0,i}}^2 + \sigma_{int}^2},
\end{equation}
where $\mathbf{p} \equiv (H_0,\alpha,\Omega_m)$ is the complete set of parameters that we want to fit, $\sigma_{\mu_{0,i}}$, $\sigma_{int}$ are, respectively,  the errors associated with the observational techniques in determining the distance moduli (includes a peculiar
contribution) and the intrinsic dispersion of SNe Ia.  The corresponding errors are reported in Astier {\it et al.} paper \cite{Astier06}.

Marginalizing our likelihood function over the
nuisance parameter, $H_0$, we obtain 
the likelihood function for the $\Omega_m - \alpha$ plane.  In order to determine the cosmological
parameters ($\Omega_m$,$\alpha$), a $\chi^{2}$ minimization
for the range of [0,1] in steps of $0.01$ has been applied. The $68.3\%$, $90.0\%$
and $95.4\%$ confidence levels are defined by the conventional
two-parameters $\chi^{2}$ levels $2.30$, $4.61$ and $6.17$,
respectively. It is very important to note that we do not consider
any prior in $\Omega_m$, as usually required by the SNe Ia test.
\begin{figure*}
\centerline{\psfig{figure=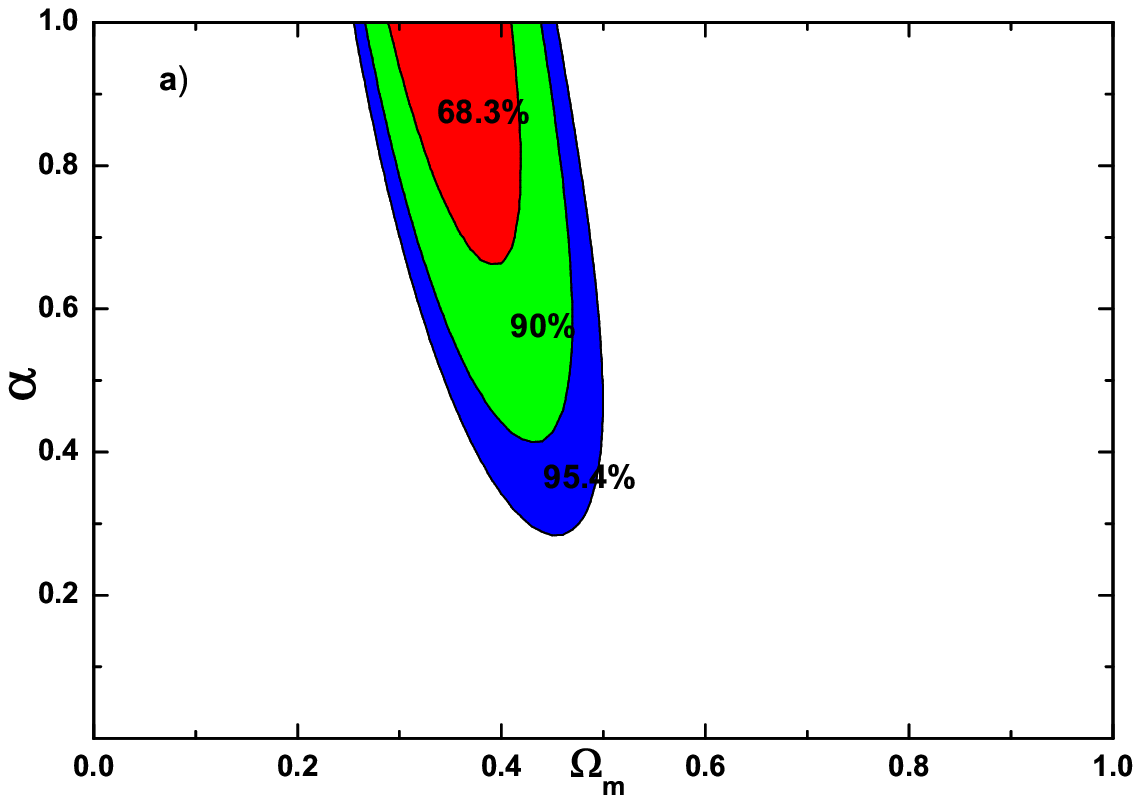,width=2.5truein,height=2.5truein}
\psfig{figure=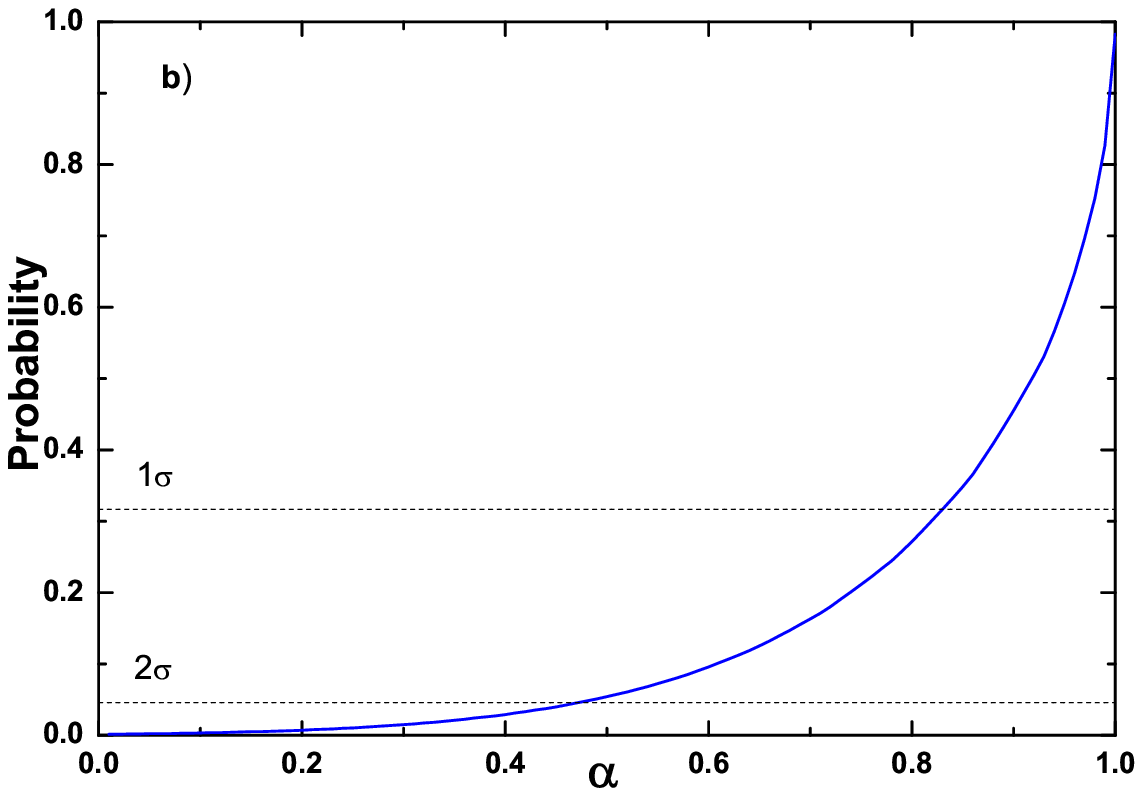,width=2.5truein,height=2.5truein}
\psfig{figure=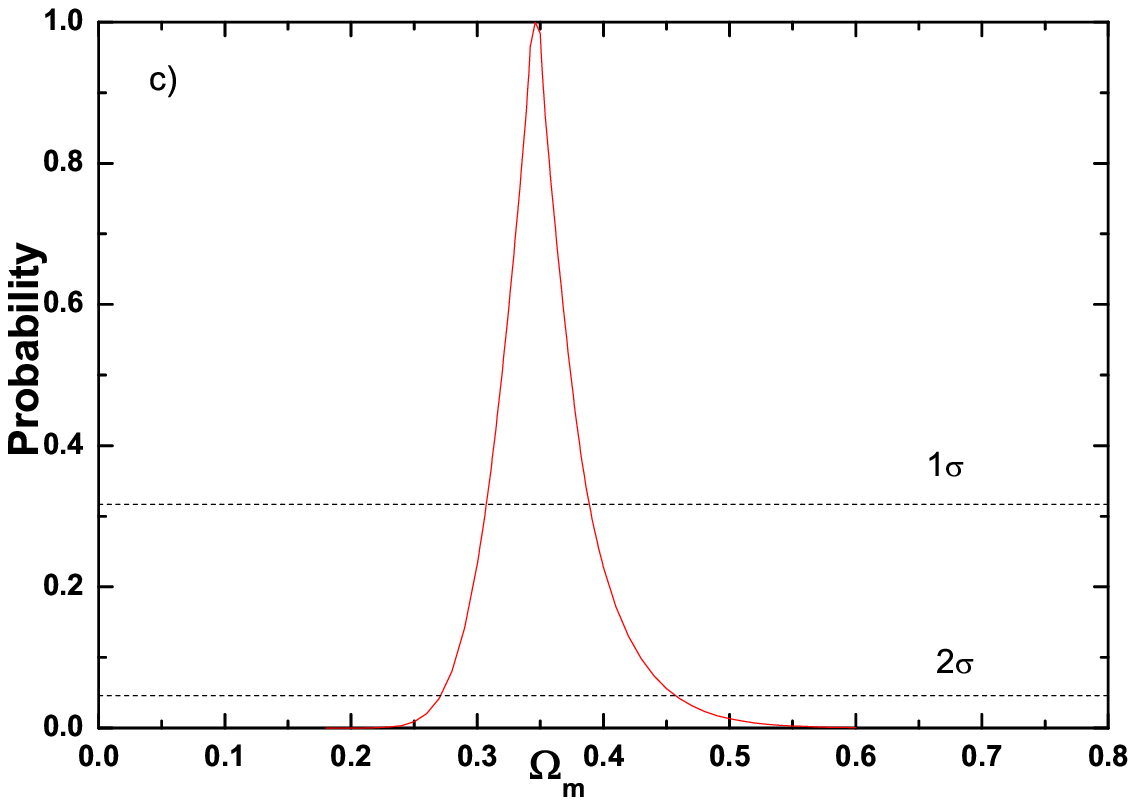,width=2.5truein,height=2.5truein}
\hskip 0.1in} \caption{{\bf{a)}} Confidence contours  on the ($\Omega_m,\alpha$) plane for  flat $\Lambda$CDM models as inferred from 182 SNe Ia measurement by Riess {\it{et al.}}
\cite{Ries07}. {\bf{b)}} The Likelihood
function for the $\alpha$ smoothness parameter. We see that at 2$\sigma$ the 
smoothness parameter is restricted on the interval ($0.42 \leq  \alpha \leq 1.0$). {\bf{c)}}
Probability of the matter density parameter. In this case
a comparatively small region is permitted $0.25 \leq \Omega_m \leq 0.44$ with 
($2\sigma$) confidence level.}
\end{figure*}
The basic results are shown in
Figs. 2a, 2b and 2c. From fig. 2a we see that all range for $\alpha$ is accepted,
while a $\Omega_m \approx 0.3\pm 0.1$ is obtained. In Fig. 3b we see the likelihood for the smoothness parameter.  The best
fit adjustment occurs for values of $\alpha=1.0$ and $\Omega_m=0.26$ with $\chi^2_{min}= 113.3$ and $\nu=113$ degrees of freedom  ($\chi^2/{\nu}=1$), thereby showing that the model provides a good fit to these data. It is also interesting that for any $\alpha$ value, we also find no evidence for a high $\Omega_m$ parameter as required by a flat Einstein-de Sitter universe ($\Omega_{\Lambda}=0$). Actually, the Einstein-de Sitter scenario has 
a very small statistical significance  $\chi^2=244.9$ ($11.5\sigma$ outside). However, since the Astier {\it et al.} data are not restrictive for the $\alpha$ parameter, let us now consider the enlarged SNeIa sample observed by the High-z Supernovae Search Team \cite{Ries07}.

\subsection{Riess et al. Sample (2007)}

The so-called \emph{gold} sample from the HZS team \cite{Ries07}
is a selection of 182 SNe Ia events distributed over the redshift
interval $0.01 \lesssim z \lesssim 1.755$, and constitutes the
compilation of the best observations made so far by them and by
the Supernova Cosmology Project events observed by
Hubble Space Telescope (HST). As before, constrains on the cosmological
parameters ($\Omega_m$,$\alpha$), are determined from a $\chi^{2}$ minimization
within  the range of [0,1] spanned by such parameters. 

In fig. 3a one can  see that $0.42 \leq  \alpha \leq 1.0$ and $0.25 \leq \Omega_m \leq 0.44$ with $90\%$ of statistical confidence.  
The best fit adjustment occurs for values of $\Omega_m= 0.33$ and $\alpha= 1$  with $\chi^2_{min}= 158.6$ and $\nu=180$ degrees of freedom the reduced ($\chi^2/{\nu}\sim 0.9$). Therefore, the model provides a very good fit to the Riess {\it et al.} sample. In Fig. 3b we see the likelihood for the smoothness parameter. 
As previously remarked, the Riess {\it et al.} data set is much more restrictive for the 
smoothness parameter than Astier {\it et al.} sample. Within $2\sigma$, the allowed range for  the $\alpha$ falls on the interval $0.42 \leq  \alpha \leq 1.0$ (cf fig 2a). In fig 3c we show the probability for the density  matter parameter. In this case
a small region is permitted $0.25 \leq \Omega_m \leq 0.44$ with of
the confidence level ($2\sigma$). In the analysis for the Einstein-de Sitter universe ($\Omega_m=1.0$, $\Omega_{\Lambda}=0.0$), the $\chi^2=255.8$ is too bad ($9.9\sigma$
c.l. outside for 1 degree of freedom), and guarantee us to exclude
this model with high confidence.

\section{Comments and Conclusions}

Cosmology is in an exciting period.  A considerable set of rather sophisticated  experiments, until a few years ago regarded as futuristic,  have now been completed with spectacular success.  The results of the first observations almost one decade ago have been confirmed what was long surmised, namely,  that most of
the matter is nonbaryonic and that we live in an accelerating expanding Universe dominated by dark energy.

In this article, we have discussed the influence of
inhomogeneities on the expansion rate of the universe. 
In particular, if the smoothness $\alpha$ parameter 
could be constrained through a statistical 
analysis  involving two large sets of SNe Ia data. 
As we have seen, in the case of the Astier {\it et al.} 
sample, the entire interval of $\alpha$ is allowed while a 
$\Omega_m \approx 0.3\pm 0.1$ is obtained. Within the existing uncertainties, these
results are consistent with the constraints obtained from angular diameter of compact radio sources with basis on the Gurvits {\it et al.} data
\cite{SL07,AL04}). Therefore, although in close agreement with rather different analysis, 
this SNe data set is uncapable to constrain the smoothness parameter. 
Actually, at this moment, the sample of Riess {\it et al.} provides a more stringent constrain with the allowed range for $\alpha$ falling on the interval $0.42 \leq  \alpha \leq 1.0$ ($2\sigma$). Basically, this occurs because the Riess {\it et al.} sample extends to appreciably higher redshifts. In general, both analysis suggest that a large range for $\alpha$ is permitted, and that the Einstein-de Sitter model is strongly excluded ($11.5\sigma$ and $9.9\sigma$, respectively).  As we have seen,  the necessity of the dark energy and a transition from deceleration to an accelerating phase is maintained even when one takes into account the clustering phenomenon.  However, at the level of the SNe Ia observations discussed here, these results suggest  that the clumpiness of matter distribution can mimic at least a small fraction of the dark energy component.

Finally, we would like to stress that measurements from SNe Ia combined with the ZKDR inhomogeneous approach adopted here may provide an independent and more rigorous cosmological test for the cosmic concordance model in the near future. In this concern, it should be very important to investigate whether the $\alpha$ parameter can be constrained using independent observations, among them: the cosmic microwave background anisotropies, the physics of galaxies clusters, Sunyaev-Zeldovich effect, time delay and statistical of gravitational lensing. Some studies along these lines will be presented in a forthcoming communication. The present results based only on the Hubble-Sandage diagram show that the Riess {\it et al.} sample more restrictive than the Astier {\it et al.} sample, thereby reinforcing the interest to observe more supernova events at higher redshifts.

\begin{acknowledgements}
We are very grateful to the organizers of the IWARA (International Workshop on Relativistic Astrophysics, Natal, Brazil) where the preliminary results of this work were presented. The authors would also like to thank L. Marassi, F. J. Jesus, R. F. L. Holanda,  and V. Busti for helpful discussions. RCS is supported by CNPq No. 15.0293/2007-0 and JVC is supported by FAPESP No. 2005/02809-5. JASL is partially supported by CNPq and FAPESP No. 04/13668-0 (Brazilian Research Agencies).
\end{acknowledgements}

\end{document}